\documentclass[twocolumn,showpacs,preprintnumbers,amsmath,amssymb]{revtex4}

\usepackage{graphicx}
\usepackage{dcolumn}
\usepackage{bm}

\begin{document}

\preprint{APS/123-QED}

\title{Decoherence induced by squeezing control errors in optical and ion trap holonomic quantum
computations}

\author{Viatcheslav I. Kuvshinov}
\email{v.kuvshinov@sosny.bas-net.by}
\affiliation{%
Joint Institute for Power and Nuclear Research, 220109 Krasina
str., 99, Minsk, Belarus. }

\author{Andrei V. Kuzmin}
\email{avkuzmin@sosny.bas-net.by}
\affiliation{%
Joint Institute for Power and Nuclear Research, 220109 Krasina
str., 99, Minsk, Belarus.
}

\date{\today}

\begin{abstract}
We study decoherence induced by stochastic squeezing control
errors considering the particular implementation of Hadamard gate
on optical and ion trap holonomic quantum computers. We
analytically obtain both the purity of the final state and the
fidelity for Hadamard gate when the control noise is modeled by
Ornstein-Uhlenbeck stochastic process. We demonstrate the purity
and the fidelity oscillations depending on the choice of the
initial superimposed state. We derive a linear formulae connecting
the gate fidelity and the purity of the final state.
\end{abstract}

\maketitle

\section{Introduction}

Holonomic quantum computations exploiting non-abelian geometrical
phases~\cite{Wilczek1} was primarily proposed in the
Ref.~\cite{HQC1} and developed further in the Ref.~\cite{HQC2}.
Many implementations of holonomic quantum computers (HQC) have
been proposed. Particularly, the realization of HQC within quantum
optics was suggested (optical HQC)~\cite{OHQC}. Laser beams in a
non-linear Kerr medium were exploited for this purpose. Two
different sets of control devices can be used in this case. The
first one considered in this paper consists of one- and two-mode
displacing and squeezing devices. The second one includes SU(2)
interferometers. As well trapped ions with the excited state
connected to a triple degenerate subspace (four level
$\Lambda$-system) can be used to implement HQC~\cite{trio}.
Another approach to HQC exploiting squeezing and displacement of
the trapped ions vibrational modes was suggested in the
Ref.~\cite{TRIOHQC}. This implementation of HQC is mathematically
similar to the first embodiment of the optical HQC~\cite{OHQC} and
thus it is also considered in this work. Particularly, expressions
for the adiabatic connection and holonomies are the same in these
cases. Another proposed implementation of HQC was the HQC with
neutral atoms in cavity QED~\cite{cavity}. The coding space was
spanned by the dark states of the atom trapped in a cavity.
Dynamics of the atom was governed by the generalized
$\Lambda$-system Hamiltonian. Mathematically similar
semiconductor-based implementation of HQC was proposed in the
Refs.~\cite{semicond}, where one-qubit gates were also realized in
the framework of the generalized $\Lambda$-system. In distinction
from the cavity model of HQC its physical implementation exploits
semiconductor excitons driven by sequences of laser
pulses~\cite{semicond}. For the two-qubit gate implementation the
bi-excitonic shift was used. The generalized $\Lambda$-system with
the different Rabi frequencies parametrization was exploited
recently for HQC implemented by Rf-SQUIDs coupled through a
microwave cavity~\cite{SQUID}. One more solid state implementation
of HQC based on Stark effect was proposed in the
Ref.~\cite{Stark}.

Let us briefly remind the main results concerning the holonomic
quantum computation. In HQC non-abelian geometric phases
(holonomies) are exploited to implement unitary transformations
over the quantum code. The later is some degenerate subspace $C^N$
spanned on eigenvectors of Hamiltonian $H_0$, which initiates the
parametric isospectral family of Hamiltonians $F = \{ H(\lambda) =
U(\lambda) H_0 U^\dag (\lambda)\}_{\lambda \in M}$. Here
$U(\lambda)$ is a unitary operator, $\lambda$ is a vector
belonging to the space of the control parameters $M$ and $N$
denotes the dimension of the degenerate computational
subspace~\cite{HQC1,HQC2}. Quantum gates are implemented when the
control parameters are adiabatically driven along the loops in the
control manifold $M$. The unitary operator mapping the initial
state vector belonging to $C^N$ into the final one has the form
$e^{i \phi} \Gamma_\gamma (A_\mu)$, where the index $\mu$
enumerates control parameters, $\lambda_\mu$ constitute vector
$\lambda$ and $\phi$ is the dynamical phase. Holonomy associated
with the loop $\gamma \in M$ is
\begin{equation}\label{holonomy}
    \Gamma_\gamma (A_\mu) = {\bf \hat{P}} \exp{\left\{ \int\limits_\gamma A_\mu d \lambda_\mu
    \right\}}.
\end{equation}
Here ${\bf \hat{P}}$ denotes the path ordering operator, $A_\mu$
is the matrix valued adiabatic connection given by the
expression~\cite{Wilczek1}:
\begin{equation}\label{connection}
    \left( A_\mu \right)_{mn} = \left< \varphi_m \right| U^\dag
    \frac{\partial}{\partial \lambda_\mu} U \left| \varphi_n
    \right> ,
\end{equation}
where $\left| \varphi_k \right>$ with $k = \overline{1,N}$ are the
eigenvectors of the Hamiltonian $H_0$ forming the basis in $C^N$.
Dynamical phase $\phi$ will be omitted bellow due to the suitable
choice of the zero energy level. We shall consider the single
subspace $C^N$ (no energy level crossings are assumed).

It is evident that the quantum gate (holonomy) performed depends
on the path passed in the control parameters space. As well it is
obvious that in real experiments it is impossible to pass the
desired loop in the control manifold without any deviations.
Errors in the assignment of the classical control parameters
$\lambda$ are unavoidable. The question about robustness of
holonomic quantum computations with respect to the control errors
has attracted a lot of attention recently. Namely, the effect of
the errors originated from the imperfect control of classical
parameters was studied for ${\bf CP}^n$ model of HQC in the
Ref.~\cite{Err1} where the control-not and Hadamard gates were
particularly considered. Berry phase for the spin $1/2$ particle
in a classical fluctuating magnetic field was considered in the
Ref.~\cite{spin12}. Approach based on the non-abelian Stokes
theorem~\cite{Stokes} was proposed in the Ref.~\cite{wePLA1}.
Namely, the general expression for the fidelity valid for
arbitrary implementation of HQC in the case of the single control
error having arbitrary size and duration was derived. Simple
approximate formulae was obtained in the small error limit.
Adiabatic dynamics of quantum system coupled to a noisy classical
control field was studied in the Ref.~\cite{Gaitan}. It was
demonstrated that stochastic phase shift arising in the
off-diagonal elements of the system's density matrix can cause
decoherence. The efficiency of Shor algorithm~\cite{Shor1} run on
a geometric quantum computer was investigated in the case when the
decoherence induced by the stochastic control errors was taken
into account. The study of the robustness of the non-abelian
holonomic quantum gates with respect to the stochastic
fluctuations of the control parameters was presented in the
Ref.~\cite{ThreeRegimes}. Three stability regimes were
discriminated in this work for the HQC model with qubits given by
polarized excitonic states controlled by laser pulses. Noise
cancellation effect for simple quantum systems was considered in
the Ref.~\cite{Solinas04}. Robustness of the parametric family of
quantum gates subjected to stochastic fluctuations of the control
parameters was studied in the Ref.~\cite{Zhu04}. Usage of the
cyclic states~\cite{cyclicSt} allowed to consider quantum gates
which could be continuously changed from dynamic gates to purely
geometric ones. It was shown that the maximum of the gate fidelity
corresponds to quantum gates with a vanishing dynamical phase.
Robust Hadamard gate implementation for optical~\cite{OHQC} and
ion trap~\cite{TRIOHQC} holonomic quantum computers was proposed
in the Ref.~\cite{wePLA2}. The cancellation of the small squeezing
control errors up to the fourth order on their magnitude was
demonstrated. Hadamard gate is one of the key elements of the main
quantum algorithms, for instance see~\cite{Shor1, Grover1}. Thus
the search for its robust implementations is of importance.

During the last few years much attention has been payed to the
study of both abelian and non-abelian geometric phases in the
presence of decoherence which is the most important limiting
factor for quantum computations. Let us briefly overview some of
these works. The abelian geometric phase of the two-level quantum
system interacting with a one and two mode quantum field subjected
to the decoherence was considered in the Ref.~\cite{DecohField}.
It was demonstrated that when the geometric phase is generated by
an adiabatic evolution the first correction due to the decoherence
of the driving quantized field for the no-jump trajectory has the
second order in the decaying rate of the field but it is not the
case for the non-adiabatic evolution. Non-abelian holonomies in
the presence of decoherence were investigated in the
Ref.~\cite{Guridi} using the quantum jump approach. The effects of
environment on a universal set of holonomic quantum gates were
analyzed. Refocusing schemes for holonomic quantum computation in
the presence of dissipation were discussed in the
Ref.~\cite{refocus}. It has been shown that non-abelian geometric
gates realized by means of refocused double-loop scheme possessed
a certain resilience against decoherence. Quantum Langevin
approach has been used to study the evolution of two-level system
with a slowly varying Hamiltonian and interacting with a quantum
environment modeled as a bath of harmonic
oscillators~\cite{HarBath}. It allowed to obtain the dissipation
time and the correction to Berry phase in the case of adiabatic
cyclic evolution. The realization of universal set of holonomic
quantum gates acting on decoherence-free subspaces has been
proposed in the Ref.~\cite{DecohFree}. It has been shown how it
can be implemented in the contexts of trapped ions and quantum
dots. The performance of holonomic quantum gates in semi-conductor
quantum dots under the effect of dissipative environment has been
studied in the Ref.~\cite{Dissipation}. It was demonstrated that
the influence of the environment modeled by the superhomic thermal
bath of harmonic oscillators could be practically suppressed. The
study of the non-adiabatic dynamics and effects of quantum noise
for the ion trap setup proposed in the Ref.~\cite{trio} has been
also done~\cite{ref29}. The optimal finite operation time was
determined. In the references mentioned above the fidelity was
used as the main measure of gate resilience.

In this paper we consider optical and ion trap implementations of
HQC proposed in the Refs.~\cite{OHQC} and~\cite{TRIOHQC}
respectively. Regarding the particular implementation of Hadamard
gate we study the decoherence induced by stochastic squeezing
control errors. Following the Ref.~\cite{spin12} we model the
random fluctuations by Ornstein-Uhlenbeck stochastic process. We
analytically obtain the final state purity and the gate fidelity
as the measures of the gate robustness with respect to the
decoherence induced by stochastic control errors. In the small
squeezing control errors limit we derive a simple formulae
connecting the purity of the final state and the gate fidelity.

\section{Hadamard gate implementation}

Optical and ion trap setups of HQC are mathematically equivalent
since the corresponding holonomies are the same (compare the
Refs.~\cite{OHQC} and \cite{TRIOHQC}). Therefore we can consider
both HQC models simultaneously.

The laser beams in the nonlinear Kerr medium are explored in order
to perform holonomic quantum computation in the framework of the
optical setup. The corresponding interaction Hamiltonian
describing a single beam in the medium is
\begin{equation}\label{HIopt}
    H_I = \hbar X a^\dag a \left( a^\dag a -1 \right),
\end{equation}
where $a$ and $a^\dag$ are the annihilation and creation operators
of the photons respectively, $X$ is a constant proportional to the
third order nonlinear susceptibility of the medium. The degenerate
computational subspace of the single qubit is spanned on the
photon Fock states $\left| 0 \right>$ and $\left| 1 \right>$. More
details one can find in the Ref.~\cite{OHQC}.

In order to implement the same holonomic quantum computational
scheme in the framework of the ion trap setup one has to deal with
the two-level trapped ion placed in the common node of two
standing electromagnetic waves with frequencies $\omega_0 -
\omega_z$ and $\omega_0 + \omega_z$ as well as being affected by
the traveling wave with the frequency $\omega_0$. Here $\omega_0$
denotes the frequency corresponding to the transition between the
two ion levels being exploited, $\omega_z$ is the frequency of the
ion's harmonic oscillations along the $z$ axis of the linear Paul
trap. The basis qubit states are $\left| g \right> \otimes \left|
0 \right> $ and $\left( \left| g \right> \otimes \left| 1 \right>
- \left| e \right> \otimes \left| 0 \right> \right) / \sqrt{2}$.
Here $\left| g \right>$ and $\left| e \right>$ are the ground and
excited internal states of the ion, $\left| 0 \right>$ and $\left|
1 \right>$ are the two lowest vibrational Fock states of the ion
in the trap. More details can be found in the Ref.~\cite{TRIOHQC}.

One-qubit gates are given as a sequence of single mode squeezing
and displacing operations~\cite{OHQC, TRIOHQC}:
\begin{equation}\label{UDS}
    U (\eta, \nu) = D(\eta) S(\nu),
\end{equation}
where
\begin{eqnarray}\label{SD}
    &&S(\nu) = \exp{\left( \nu a^{\dag 2} - \nu^* a^2
    \right)}, \nonumber \\
    &&D(\eta) = \exp{\left( \eta a^\dag - \eta^* a \right)}
\end{eqnarray}
denote single mode squeezing and displacing operators
respectively, $\nu = r_1 e^{i \theta_1}$ and $\eta = x +iy$ are
corresponding complex control parameters, $a$ and $a^\dag$ are
annihilation and creation operators. The asterix denotes complex
conjugation. The expressions for the adiabatic connection and the
curvature tensor can be found in the Refs.~\cite{OHQC, TRIOHQC}.
Following our previous Letter~\cite{wePLA2} we consider Hadamard
gate
\begin{equation}\label{Hadamard0}
    H_0 = \frac{1}{\sqrt{2}} \left(%
\begin{array}{cc}
  1 & 1 \\
  1 & -1 \\
\end{array}%
\right)
\end{equation}
implemented when two rectangular loops belonging to the planes
$(x,r_1)\left|_{\theta_1 = 0} \right.$ and
$(y,r_1)\left|_{\theta_1 = 0} \right.$ are passed. Namely,
\begin{equation}\label{HGG}
    -iH_0 = \Gamma (C_{II})\left|_{\Sigma_{II} = \pi /2} \right. \Gamma
    (C_I) \left|_{\Sigma_I = \pi /4} \right. ,
\end{equation}
where the holonomies are
\begin{eqnarray}\label{Gammas}
    \Gamma (C_I) = \exp{\left( -i\sigma_y \Sigma_I \right)}, \quad
    \Sigma_I = \int\limits_{S(C_I)} dx dr_1 2 e^{-2r_1}, \nonumber
    \\
    \Gamma (C_{II}) = \exp{\left( - i \sigma_x \Sigma_{II}
    \right)}, \quad \Sigma_{II} = \int\limits_{S(C_{II})} dy dr_1
    2 e^{2r_1},
\end{eqnarray}
and $S(C_{I,II})$ are the regions in the planes
$(x,r_1)\left|_{\theta_1 = 0} \right.$ and
$(y,r_1)\left|_{\theta_1 = 0} \right.$ enclosed by the rectangular
loops $C_I$ and $C_{II}$ respectively. The sides of the rectangles
$C_I$ and $C_{II}$ are parallel to the coordinate axes. For the
loop $C_I$ these sides are given by the lines $r_1 = 0$, $x =
b_x$, $r_1 = d_x$ and $x= a_x$, where the length of the
rectangle's sides parallel to the $x$ axis is $l_x = b_x - a_x$.
In the Ref.~\cite{wePLA2} it was shown that
\begin{equation}\label{dx}
    d_x = - \frac{1}{2} \ln{\left( 1 - \frac{\pi}{4l_x} \right)},
    \quad l_x > \frac{\pi}{4} .
\end{equation}
In the same way the rectangle $C_{II}$ is composed of the lines
$r_1 = 0$, $y = b_y$, $r_1 = d_y$ and $y = a_y$,
where~\cite{wePLA2}:
\begin{equation}\label{dy}
    d_y = \frac{1}{2} \ln{\left( 1 + \frac{\pi}{2l_y} \right)}.
    \quad l_y = b_y - a_y.
\end{equation}
Proposed Hadamard gate implementation is not a unique one. The
same gate can be realized by passing another loops in the control
manifold. Our choice is motivated by the simplicity of the loops.

\section{Decoherence induced by stochastic squeezing control errors}

We restrict ourselves by the consideration of the squeezing
control errors only. Moreover, we can neglect the fluctuations of
the squeezing control parameter when $r_1 = 0$. Thus to take into
account random squeezing control errors we have to replace $d_x$
by $d_x + \delta r_x (x)$ and $d_y$ by $d_y + \delta r_y (y)$,
where $\delta r_x (x)$ and $\delta r_y (y)$ are two independent
Ornstein-Uhlenbeck stochastic processes. Making this substitution
into the Eqs.~(\ref{Gammas}) instead of the formulae~(\ref{HGG})
we obtain the following expression for the perturbed Hadamard
gate, see also~\cite{wePLA2}:
\begin{eqnarray}\label{PertHadamard}
    -i H = - \frac{1}{\sqrt{2}} \left( \cos{\alpha} - \sin{\alpha}
    \right) \left( \sin{\beta} + i \sigma_x \cos{\beta} \right) -
    \nonumber \\
    - \frac{i}{\sqrt{2}} \left( \cos{\alpha} + \sin{\alpha}
    \right) \left( \sigma_z \cos{\beta} - \sigma_y \sin{\beta}
    \right),
\end{eqnarray}
where
\begin{eqnarray}\label{albe}
    \alpha = e^{-2d_x} \int\limits_{a_x}^{b_x} dx \left( 1 - e^{-2\delta r_x}
    \right), \nonumber \\
    \beta = e^{2d_y} \int\limits_{a_y}^{b_y} dy \left( e^{2 \delta r_y} - 1
    \right).
\end{eqnarray}
Let the qubit initially to be in the pure superimposed state
$\left| \psi_0 \right> = c_0 \left| 0 \right> + c_1 \left| 1
\right>$, where amplitudes $c_0$ and $c_1$ obey the normalization
constrain $\left| c_0 \right|^2 + \left| c_1 \right|^2 = 1$. For
the fixed noise realization the final qubit state will be pure.
However, it will differ from the desired one. In the real
experiment we do not follow the random fluctuations of the control
parameters (nevertheless {\it in principle} we can do it). In this
situation quantum mechanics prescribes us to describe the final
state of the system by the density matrix and represent the state
as a mixture of all possible final states weighted with the
probabilities of the corresponding noise realizations. Following
this strategy we find the density matrix of the final state for a
given noise implementation and than average over the squeezing
control parameter fluctuations when the later are modeled by the
two independent Ornstein-Uhlenbeck stochastic processes.

Thus, for the density operator $\rho \equiv H \left| \psi_0
\right> \left< \psi_0 \right| H^\dag$ we obtain the following
matrix elements:
\begin{widetext}
\begin{eqnarray}\label{tilderho}
    \left< 0 \right| \rho \left| 0 \right>  =
    \frac{1}{2} + \frac{1}{2} \left( \left| c_0 \right|^2 - \left| c_1 \right|^2 \right) \cos{2\beta} \cos{2\gamma}
    - \frac{1}{2} \left( c_0 c_1^* + c_0^* c_1 \right) \cos{2\beta} \sin{2\gamma}
    - \frac{i}{2} \left(
    c_0 c_1^* - c_0^* c_1 \right)\sin{2\beta} ,
    \nonumber \\
    \left< 0 \right| \rho \left| 1 \right> = \frac{1}{2} \left(
    \left| c_1 \right|^2 - \left| c_0 \right|^2 \right)
    \sin{2\gamma}
     + \frac{i}{2} \left( \left| c_0 \right|^2 - \left| c_1 \right|^2
    \right) \sin{2\beta} \cos{2\gamma}
    - \frac{1}{2} \left( c_0 c_1^* + c_0^* c_1 \right) \left(
    \cos{2\gamma} + i\sin{2 \beta} \sin{2\gamma} \right) \nonumber
    \\
    - \frac{1}{2} \left( c_0 c_1^* - c_0^* c_1 \right)
    \cos{2\beta}, \nonumber \\
    \left< 1 \right| \rho \left| 0 \right> = \left< 0 \right| \rho \left| 1
    \right>^*, \quad
    \left< 1 \right| \rho \left| 1 \right> = 1 - \left< 0 \right| \rho \left| 0
    \right>.
\end{eqnarray}
\end{widetext}
Here the introduced parameter $\gamma$ is defined as $\gamma =
\alpha - \pi /4$ and the asterix denotes the complex conjugate
quantities.

We assume that the noise $\delta r_x$ has variance
$\tilde{\sigma}_x$ and a lorentzian spectrum with the bandwidth
$\Gamma_x$. The fluctuations $\delta r_y$ have the variance
$\tilde{\sigma}_y$ and bandwidth $\Gamma_y$. Using the
Eqs.~(\ref{albe})-(\ref{tilderho}) and the properties of
Ornstein-Uhlenbeck stochastic process (see Ref.~\cite{Kaiser}):
\begin{eqnarray}\label{OUhProc}
    \overline{\delta r_{x,y} (t_1) \delta r_{x,y} (t_2)} =
    \tilde{\sigma}_{x,y} e^{- \Gamma_{x,y} \left| t_1 - t_2
    \right|}, \nonumber \\
    \overline{\delta r_{x,y}^2} = \tilde{\sigma}_{x,y} , \quad \overline{\delta r_{x,y}} =
    0,
\end{eqnarray}
we average the density matrix $\rho$ over the stochastic
fluctuations of the squeezing control parameters $\delta r_x$ and
$\delta r_y$. The line over the random quantities means the
averaging operation. We assume that $\delta r_{x,y} \ll 1$ and
restricted ourselves by the first non-vanishing terms depending on
$\delta r_x$ or $\delta r_y$. As the result of straightforward but
a bit lengthy calculations we analytically obtain the elements of
the averaged density matrix $\overline{\rho}$:
\begin{widetext}
\begin{eqnarray}\label{rhoj}
    \left< 0 \right| \overline{\rho} \left| 0 \right> =
    \frac{1}{2} \left( 1 + c_0 c_1^* + c_0^* c_1 \right) - 2 \left(
    \left| c_0 \right|^2 - \left| c_1 \right|^2 \right)
    \tilde{\sigma}_x l_x e^{-2 d_x} - 2i \left( c_0 c_1^* - c_0^* c_1
    \right) \tilde{\sigma}_y l_y e^{2 d_y} \nonumber \\
    - \left( c_0 c_1^* + c_0^* c_1 \right) \left[ \frac{8 \tilde{\sigma}_x}{\Gamma_x} {\cal F}_x + \frac{8 \tilde{\sigma}_y}{\Gamma_y}
    {\cal F}_y
    \right], \nonumber \\
    \left< 0 \right| \overline{\rho} \left| 1 \right> =
    \frac{1}{2} \left( c_0 + c_1 \right) \left( c_0^* - c_1^*
    \right) + 2 \left( c_0 c_1^* + c_0^* c_1 \right) \left(
    \tilde{\sigma}_x l_x e^{- 2 d_x} + i \tilde{\sigma}_y l_y e^{2 d_y}
    \right) \nonumber \\
    - \frac{8 \tilde{\sigma}_x}{\Gamma_x} \left( \left| c_0 \right|^2 - \left| c_1 \right|^2
    \right){\cal F}_x + \frac{8 \tilde{\sigma}_y}{\Gamma_y} \left( c_0 c_1^* - c_0^* c_1
    \right) {\cal F}_y, \nonumber \\
    \left< 1 \right| \overline{\rho} \left| 0 \right> = \left< 0 \right| \overline{\rho} \left| 1
    \right>^* , \quad \left< 1 \right| \overline{\rho} \left| 1
    \right> = 1- \left< 0 \right| \overline{\rho} \left| 0
    \right>.
\end{eqnarray}
\end{widetext}
Here we introduced the following denotations:
\begin{eqnarray}\label{FF}
    {\cal F}_x = e^{-4 d_x}
    \left( l_x - \frac{1 - e^{-\Gamma_x l_x}}{\Gamma_x} \right),
    \nonumber \\
    {\cal F}_y = e^{4 d_y}
    \left( l_y - \frac{1 - e^{-\Gamma_y l_y}}{\Gamma_y} \right).
\end{eqnarray}

In order to quantify decoherence strength we exploit the purity of
the final state. It is defined as the trace of the squared density
matrix. Purity equals to $1$ for pure states and less than $1$
overwise. Using the Eqs.~(\ref{tilderho}) it is easy to check that
for a fixed noise realization the purity $I_0 = tr \rho^2$ equals
the unity as it should be. Using the Eqs.~(\ref{rhoj}) we obtain
the purity of the final state in the case of the stochastic
squeezing control errors. The result of the lengthy but
straightforward calculations is
\begin{eqnarray}\label{I}
    I = tr \overline{\rho}^2 = 1 - \frac{64 \tilde{\sigma}_y}{\Gamma_y} {\cal F}_y \left| c_0
    \right|^2 \left| c_1 \right|^2 \nonumber \\ - \frac{16
    \tilde{\sigma}_x}{\Gamma_x} {\cal F}_x \left( c_0^2 + c_1^2
    \right) \left( c_0^{*2} + c_1^{*2} \right),
\end{eqnarray}
Thus we see that the final state purity $I<1$ and the stochastic
squeezing control errors induce decoherence and lead the final
state to be a mixture of pure states. We can simplify the
expression~(\ref{I}) if we exploit the following parametrization
for the amplitudes $c_0$ and $c_1$:
\begin{equation}\label{ccphi}
    c_0 = e^{i \xi} \cos{\varphi} , \quad c_1 = e^{i \chi}
    \sin{\varphi},
\end{equation}
and assume that $\xi -\chi = \pi n $, where $n$ is an integer. In
this rather general case formulae~(\ref{I}) reduces to
\begin{equation}\label{Ia}
    I = 1 - \frac{16 \tilde{\sigma}_x}{\Gamma_x} {\cal F}_x -
    \frac{16 \tilde{\sigma}_y}{\Gamma_y} {\cal F}_y
    \sin^2{2\varphi} .
\end{equation}
We see that the final state purity oscillates depending on the
choice of the initial qubit state. The purity decreases when the
initial qubit state is a superposition of the basis states. For
example, it has its minimum when $\left| \psi_0 \right> = \left(
\left| 0 \right> \pm \left| 1 \right> \right)/\sqrt{2}$. In
contrary, the purity has its maximum if the initial qubit state
$\left| \psi_0 \right>$ is proportional to one of the basis states
$\left| 0 \right>$ or $\left| 1 \right>$. In this case the errors
made in the $(y,r_1)\left|_{\theta_1 = 0} \right.$ plane are
eliminated.

\section{Hadamard gate fidelity}

Now we obtain the fidelity of the non-ideal Hadamard gate. In the
case when there are no control errors ($\delta r_x = \delta r_y =
0$) the density matrix $\rho^{(0)}$ of the final (pure) state has
the following form:
\begin{widetext}
\begin{eqnarray}\label{rho0}
    \rho^{(0)} \equiv H_0 \left| \psi_0 \right>\left< \psi_0
    \right| H_0 =
    \frac{1}{2}\left(%
\begin{array}{cc}
  1 + c_0^* c_1 + c_0 c_1^* & \left| c_0 \right|^2 - \left| c_1 \right|^2 +c_0^* c_1 - c_0 c_1^* \\
  \left| c_0 \right|^2 - \left| c_1 \right|^2 + c_0 c_1^* - c_0^* c_1 & 1 - c_0^* c_1 - c_0 c_1^* \\
\end{array}%
\right).
\end{eqnarray}
\end{widetext}
The non-ideal Hadamard gate fidelity $F \equiv tr (\rho^{(0)}
\overline{\rho})$ under the same assumptions as in the
Eq.~(\ref{rhoj}) is given by the expression
\begin{equation}\label{F}
    F = 1 - \frac{32 \tilde{\sigma}_y}{\Gamma_y} {\cal F}_y \left|
    c_0 \right|^2 \left| c_1 \right|^2 - \frac{8
    \tilde{\sigma}_x}{\Gamma_x} {\cal F}_x \left( c_0^2 + c_1^2
    \right) \left( c_0^{*2} + c_1^{*2} \right).
\end{equation}
From this expression we can conclude that when the initial state
vector is proportional either to $\left| 0 \right>$ or $\left| 1
\right>$ the contribution of the control errors made in the
$(y,r_1)\left|_{\theta_1 = 0} \right.$ plane can be neglected at
the accepted approximation degree. In our previous
work~\cite{wePLA2} the fidelity was defined as $f = \sqrt{F}$ and
the condition $\left|
    c_0 \right|^2 \left| c_1 \right|^2 = 0$ was assumed. In the limit $(\Gamma_x l_x)^{-1} \to 0$, when
the fluctuations average out, from the Eq.~(\ref{F}) we obtain
that $1-f \sim \tilde{\sigma}_x^2$. Thus our previous
result~\cite{wePLA2} concerning the cancellation of the squeezing
control errors up to the fourth order on their magnitude is
reproduced as it should be (remind that $\tilde{\sigma}_x$ has the
order of $(\delta r_x)^2$).

We can simplify the expression~(\ref{F}) exploiting the
parametrization~(\ref{ccphi}) for the amplitudes $c_0$ and $c_1$
and assuming $\xi -\chi = \pi n $, where $n$ is an integer. Under
these assumptions the gate fidelity is given by
\begin{equation}\label{Fa}
    F = 1 - \frac{8 \tilde{\sigma}_x}{\Gamma_x} {\cal F}_x -
    \frac{8 \tilde{\sigma}_y}{\Gamma_y} {\cal F}_y \sin^2{2
    \varphi}.
\end{equation}
It is evident that the gate fidelity oscillates depending on the
choice of the initial qubit state. It is less for the superimposed
initial states than for the basis ones. Namely, the fidelity has
its maximum when the initial qubit state $\left| \psi_0 \right>$
is proportional either to $\left| 0 \right>$ or $\left| 1
\right>$. In this case the errors made in the
$(y,r_1)\left|_{\theta_1 = 0} \right.$ plane are eliminated. As
well the fidelity has its minimum when the initial state vector is
equal to $\left( \left| 0 \right> \pm \left| 1 \right>
\right)/\sqrt{2}$. Moreover from Eqs.~(\ref{I}) and (\ref{F}) we
find a simple linear formulae connecting the purity of the final
state and the gate fidelity:
\begin{equation}\label{FI}
    I = 2F -1.
\end{equation}
This expression demonstrates the close connection between these
quantities defining the decoherence strength and the gate
stability.

\section{Conclusions}

We considered optical and ion trap HQC proposed the
Refs.~\cite{OHQC} and \cite{TRIOHQC} respectively. Regarding the
particular implementation of Hadamard gate we have studied
decoherence induced by stochastic squeezing control errors.
Ornstein-Uhlenbeck stochastic process was exploited to model
random fluctuations of the squeezing control parameter. We have
analytically obtained the purity of the final qubit state and
calculated the fidelity of the non-ideal Hadamard gate. It was
shown that the stochastic squeezing control errors reduce the
final state into a mixture of pure states and, thus, induce
decoherence. We have shown that the final state purity oscillates
depending on the choice of the initial qubit state. The purity
decreases when the initial qubit state is a superposition of the
basis states. For example, it has its minimum when the initial
state vector is equal to $\left( \left| 0 \right> \pm \left| 1
\right> \right)/\sqrt{2}$. In contrary, the purity has its maximum
when the initial qubit state is proportional to one of the basis
states $\left| 0 \right>$ or $\left| 1 \right>$. In this case the
control errors made in the $(y,r_1)\left|_{\theta_1 = 0} \right.$
plane are eliminated. The same conclusions can be made for the
gate fidelity. Simple linear formulae connecting the gate fidelity
and the purity of the final state was derived.

\end{document}